\documentstyle[aps,prl,epsfig,floats]{revtex} 
\draft

\begin{document}
\twocolumn[\hsize\textwidth\columnwidth\hsize\csname
@twocolumnfalse\endcsname 


\title{
Scalars from Top-condensation Models at Hadron Colliders
}

\author{Gustavo Burdman\cite{email} \\ \mbox{}} 
\address{
Department of Physics, 
University of Wisconsin,\\ Madison WI 53706.
} 


\maketitle
             
\begin{abstract}
We study the production and decay  of neutral scalars and pseudo-scalars 
at hadron colliders, in theories where the top-quark mass is the result 
of a $t\bar t$ condensate. 
We show that the dominant decay channel for masses below the $t\bar t$ threshold
is the flavor changing mode $tc$. This is a consequence of the 
non-universal nature of the underlying interactions
in all top-condensation models and provides a model-independent signature
of these scenarios.
We show that an upgraded Tevatron is sensitive to a sizeable region of the 
interesting parameter space 
and that the LHC will  highly constrain these models through this
flavor violating channel. 
\end{abstract}

\pacs{PACS numbers: 12.60.Fr, 14.80.Cp, 12.60.Rc ~~~~~~~~~~~~~~~~~~~~~~~~~~~~~
~~~~~~~~~~~~~~~~MADPH-99-1116}

\vskip2.0pc]


The generation of a large fermion mass such as $m_t$ is 
a difficult problem in  theories of dynamical 
Electroweak Symmetry Breaking (ESB). The idea of the top quark mass
as a ``constituent'', dynamical mass generated by the presence of 
a condensate $\langle\bar t t\rangle$ addresses this problem, providing at the same
time a source of dynamical ESB not requiring large amounts of new matter. 
In the original top-condensation standard model~\cite{topcond}, 
the formation of the 
$\langle\bar t t\rangle$  condensate is fully responsible for the masses 
of the standard model (SM)
gauge bosons as well as for the dynamical generation of $m_t$. 
If the scale of the interaction driving the condensation is $\Lambda$, then at lower 
energies  there is a scalar doublet, the top-Higgs, which acquires a vacuum expectation 
value (VEV). Just as in the SM, the Nambu-Goldstone bosons (NGB)
are eaten by the $W$ and $Z$, leaving a neutral, CP-even scalar particle in the 
spectrum. 

Although the original top condensation model fails to provide a natural
picture of ESB, the idea that a $\langle t\bar t\rangle$ condensate is 
responsible for the
large ``constituent'' top quark mass is still appealing. The Pagels-Stokar formula 
relating the NGB decay constant to the cutoff scale $\Lambda$ and the dynamical 
top quark 
mass is~\cite{pagels}
\begin{equation}
f_{\pi_t}^2\simeq\frac{N_c}{8\pi^2}m_t^2\ln{\frac{\Lambda^2}{m_t^2}}, 
\label{pagstok}
\end{equation}
where $N_c$ is the number of colors. From Eq.~(\ref{pagstok}) it can be seen that 
obtaining the correct top quark mass {\em and} $f_{\pi_t}=v\simeq 246$~GeV 
forces the cutoff scale to be $\Lambda\simeq 10^{15}$~GeV. This results in the need 
of extreme fine tuning of the coupling of the underlying gauge interaction (Topcolor), 
reproducing the naturalness problem of the SM we set out to solve in the first place. 
On the other hand, if $\Lambda\simeq {\cal O}(1)$~TeV we obtain 
$f_{\pi_t}\approx (70-80)$~GeV, 
not enough to fully break the electroweak symmetry. Thus, one can imagine that 
there is an additional source of ESB such that this deficit is covered. This is the
case in the Topcolor Assisted Technicolor (TATC) scenario proposed by Hill~\cite{tatc}, 
where Technicolor gives most of the $W$ and $Z$ masses. In addition, in this scenario
Extended Technicolor (ETC) is invoked to generate fermion masses up to ${\cal O}(1)$~GeV,
leaving the burden of most of $m_t$ to Topcolor. 

In general, whatever the model invoked 
to supplement Topcolor in generating $v$, it results in an additional set of 
NGBs. A triplet of these new NGBs mixes with the one resulting from the breaking of 
the top quark chiral symmetry: one set is absorbed by the $W$ and the $Z$ and the 
other one, the so called top-pions, remains in the spectrum. In the TATC scenario the 
top-pions acquire
masses in the range $m_{\pi_t}\simeq(100-300)~$GeV. 
The neutral CP-even state analogous to the $\sigma$ particle in 
low energy QCD, the top-Higgs, is a $t\bar t$ bound state and its mass can be 
estimated in the Nambu--Jona-Lasinio (NJL) model in the large $N_c$ approximation, to be 
\begin{equation}
m_h\simeq 2m_t\label{mthiggs}.
\end{equation}
This estimate is rather crude and it should be taken as a rough indication of where 
the top-Higgs mass could be.  
Masses well below the $t\bar t$ threshold 
are quite possible and occur in a variety of cases~\cite{cdgh}. 

The search for scalars from top-condensation models has been previously considered 
in the literature for the case of the charged top-pions from Topcolor~\cite{yuan}.
In this paper, we will examine the prospects for the observation of the 
top-Higgs $h_t$ and 
the neutral top-pion $\pi_t^0$ at hadron colliders. 
Most viable models of top condensation feature additional scalars. Here we 
concentrate on those
that couple to $t_R$ since this endows them with a large Yukawa coupling 
\begin{equation}
r\equiv\frac{m_t}{f_{\pi_t}}\label{yukawa}.
\end{equation}
This implies that the cross sections in 
production mechanisms governed by the top-Yukawa coupling are 
enhanced by 
a factor of $r^2\approx 10$ 
with respect to the equivalent SM Higgs production processes~\cite{snow}. 
Such is the case with gluon-gluon fusion, as well as 
the emission from a top quark line. 
On the other hand, the associated production with electroweak gauge bosons, the 
search channel for the SM Higgs boson at LEP and at the Tevatron, is {\em suppressed}
by the factor $r^2$ in the case of the top-Higgs production, a reflection of the fact
that only a fraction of $M_W$ and $M_Z$ come from the top condensation mechanism. 

Just as for the SM Higgs boson, the dominant production mechanism at hadron 
colliders is through gluon-gluon fusion. 
%
%
%
%
%
The fact that the cross section
is $r^2$  times larger than in the SM is of little help if the $h_t$ and $\pi^0_t$
decay modes
are subjected to the same insurmountable backgrounds. 
A crucial observation is that the underlying interactions in top condensation models 
are non-universal and therefore do not posses a Glashow-Illiopoulos-Maiani (GIM)
mechanism. This is an essential feature of these models due to the need to single 
out the top quark for condensation. For instance in TATC models the Topcolor gauge 
interactions are $SU(3)_1\times SU(3)_2$, which breaks down to the standard $SU(3)_c$ 
of $QCD$ at the TeV scale. Here the $SU(3)_1$ couples strongly to the third generation
quarks and becomes chiral-critical giving rise to $\langle t \bar t\rangle
\not =0$. These non-universal gauge interactions result in Flavor Changing Neutral
Current (FCNC) vertices of the heavy gauge bosons, the top-gluons, when one writes
the interactions in the quark mass eigen-basis. 
As long as the top-gluon masses are in 
the several hundred ~GeV range and above, 
conflicts with low energy data are avoided~\cite{bbhk}. 
Furthermore, the neutral scalar sector of Topcolor theories exhibits the 
same FCNC vertices. For instance, the off-diagonal top-Higgs couplings
take the form
\begin{eqnarray}
\frac{m_t}{f_{\pi_t}}h_t\left\{U^{(tt)*}_L U^{(tc)}_R\bar t_Lc_R 
+ U^{(tc)*}_L U^{(tt)}_R \bar c_L t_R \right. \nonumber \\
\left. U^{(tt)*}_L U^{(tu)}_R\bar t_Lu_R 
+ U^{(tu)*}_L U^{(tt)}_R \bar u_L t_R 
\right\}.
\label{htod}
\end{eqnarray}
Here $U_L$ and $U_R$ are the left and right-handed up quark rotation matrices
that diagonalize the up quark mass matrix. The Cabibbo-Kobayashi-Maskawa (CKM)
matrix is given by $V_{CKM}=U^\dagger_L D_L$, where $D_L$ is the analogous matrix in 
the left-handed down quark sector. Thus, although the off-diagonal elements of 
$U_L$ and $U_R$ are not parameters measured independently, our knowledge of $V_{CKM}$
and the quark masses allows us to estimate the most natural region of their parameter 
space.                                       
A simple assumption is that the diagonal elements of $U_{L,R}$ and $D_{L,R}$
are close to unity and 
that they are increasingly suppressed as one moves away from the diagonal. 
Independently of the nature and size of the off-diagonal suppression, this 
gives a prescription to estimate how {\em large} the elements of these matrices
can be without the mediation of unnatural cancellations. 
Then, for instance, we have 
\begin{eqnarray}
V_{cb}&=& U^{(uc)*}_L D^{(db)}_L + U^{(cc)*}_L D^{(sb)}_L
+ U^{(ct)*}_L D^{(bb)}_L \nonumber\\
&\simeq& D^{(sb)}_L + U^{(ct)*}_L \simeq {\cal O}(\lambda^2) ,
\label{vcb}
\end{eqnarray}
and 
\begin{eqnarray}
V_{ub}&=& U^{(uu)*}_L D^{(db)}_L + U^{(cu)*}_L D^{(sb)}_L
+ U^{(tu)*}_L D^{(bb)}_L \nonumber\\
&\simeq& D^{(db)}_L + U^{(cu)*}_L D^{(sb)}_L
+ U^{(tu)*}_L \simeq {\cal O}(\lambda^3) ,
\label{vub}
\end{eqnarray}
where $\lambda\simeq 0.22$ is the Cabibbo angle in the Wolfenstein parametrization
of $V_{CKM}$. 
Then it is natural to assume $U^{(tc)}_L\simeq {\cal O}(\lambda^2)$ and that 
the $\bar u_L t_R$ coupling in Eq.~(\ref{htod}) can be neglected. 
On the other hand, there are no similar constraints 
on the off-diagonal elements of $U_R$ unless something is known 
about $M_U$, the mass matrix in the weak eigen-basis. 
For instance, it was shown in \cite{yuan} that with the 
dynamically generated triangular
textures proposed in \cite{bbhk}, and assuming that the dynamical top quark mass
makes up $99\%$ of $m_t$, one obtains the rather loose bound $U^{tc}_R<0.15$.
Other, more specific realizations of the TATC scenario~\cite{klane} suggest much smaller 
values of $U^{(tc)}_L$, although still allow large values for $U^{(tc)}_R$. 
Low energy constraints such as $D^0-\bar{D^0}$ mixing
can be easily accommodated~\cite{bbhk} 
even in this naive picture.   
Here we will treat these couplings as free parameters. The relevant 
quantity for our analysis is $U_{tc}\equiv \sqrt{|U^{(tc)}_L|^2 + |U^{(tc)}_R|^2}$. 

\begin{figure}[h]
\leavevmode
\centering
\epsfig{file=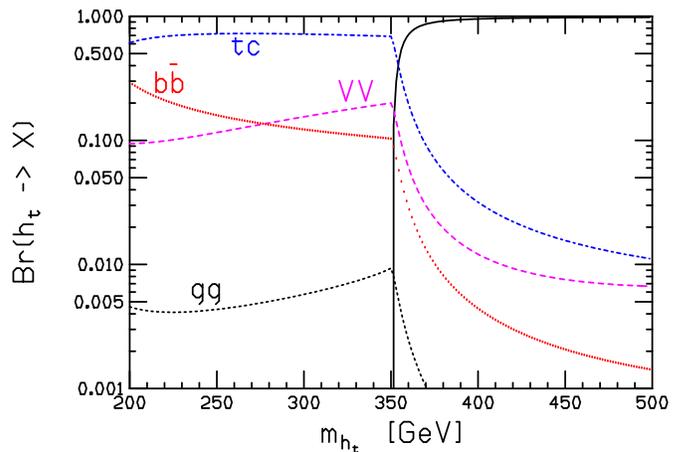,width=6cm,height=8.8cm,angle=90}
\caption{The top-Higgs branching fractions as a function of its mass.
The solid line is the $t\bar t$ decay mode, the dashed line is $VV=(WW+ZZ)$, the
the dot-dash line corresponds to $h_t\to (t\bar c+\bar tc)$, the upper and lower dotted lines 
correspond to 
$b\bar b$ and $gg$ respectively.}

\label{brs}
\end{figure}

In Fig.~\ref{brs} we plot the branching ratios of the top-Higgs 
vs. its mass. 
For the $h_t\to tc$ decay mode we use $U_{tc}=0.05$, a rather conservative value closer 
to (\ref{vcb}) than to the $U^{(tc)}_R$ upper bound. 
As mentioned above, the $WW$ and $ZZ$ are suppressed by $r^2$
with respect to the
SM case. The $b\bar b$ mode is induced by instanton effects, which we assumed 
to be responsible for about $80\%$ of the b-quark mass~\cite{tatc}.  
We see that below the $t\bar t$ threshold the dominant decay mode is the top-charm
channel with typical branching ratios of about $70\%$. 
Thus, for $m_{h_t}<2m_t$, anomalous single top production in association with a
charm jet could provide a clean signal for the discovery of the top-Higgs 
at the Fermilab Tevatron. Almost identical conclusions apply to the top-pions, 
where the only difference is the absence of the gauge boson decay channels, 
which causes the branching ratio to $tc$ to be slightly higher.
Top-pions with masses below the $t\bar t$ threshold cause large 
negative deviations in the $Z->b\bar b$ vertex~\cite{bk}
and unless some other contribution cancels them, they are in conflict with the
current measurement of $R_b$ from experiments at the $Z$ pole.
On the other hand, if such cancellations exist the top-pions could 
lie below $2m_t$ and be observed in the $\pi_t^0\to tc$ channel as well.

\begin{figure}[h]
\leavevmode
\centering
\epsfig{file=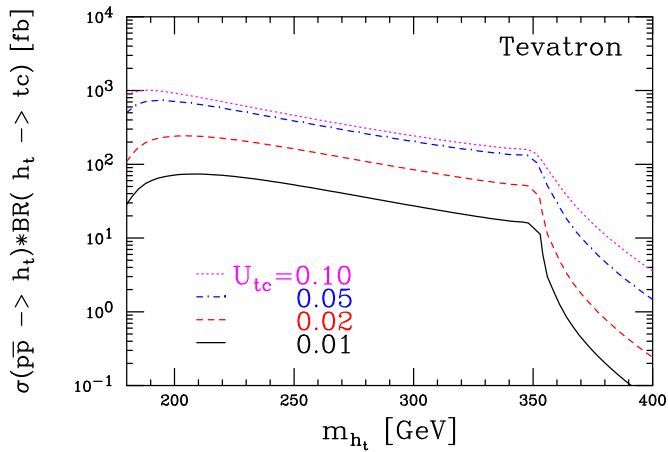,width=6cm,height=8.8cm,angle=90}
\caption{
The top-Higgs production cross section via gluon fusion times its branching ratio to 
$(t\bar c + \bar t c)$ at $\sqrt{s}=2~$TeV, for representative values of the 
parameter $U_{tc}$.
}
\label{tev}
\end{figure}

\begin{figure}[h]
\leavevmode
\centering
\epsfig{file=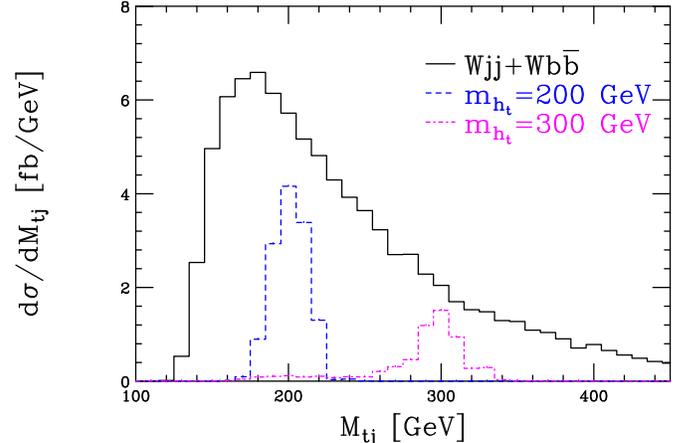,width=6cm,height=8.8cm,angle=90}
\caption{
The mass distribution for the top-charm jet system 
for $m_{h_t}=200$ and $300$~GeV, $U_{tc}=0.05$, and 
the $Wjj$ and $Wb\bar b$ backgrounds for a standard set of cuts and 
$\sqrt{s}=2~$TeV. 
We assumed $60\%$ b-tagging efficiency and  
$0.5\%$ misidentification probability of a light jet as a b jet.
}
\label{back}
\end{figure}

In Fig.~\ref{tev} we plot the top-Higgs production cross section via gluon fusion
times its branching
ratio to $(t\bar c + \bar t c)$ for various values of the parameter 
$U_{tc}$. 
For instance for values as small as $U_{tc}=0.02$  
a few hundred events are produced at the Tevatron in Run~II.  
Thus the signal is expected to be sizeable even 
after assuming $60\%$ $b$-tagging efficiency and the $e,\mu$ decay modes
of the $W$. This is more so since the signal 
presents a narrow peak in the top-jet invariant mass distribution. 
The top-Higgs total width is 
\begin{equation}
\Gamma_{h_t}<7{\rm ~GeV}.
\end{equation}
Thus, the top-charm mass distribution will present a narrow resonance 
which should be observable above the smooth background.
The main backgrounds for this process are $Wjj$, with one of the light jets 
tagged as a $b$-jet; $Wb\bar b$, where one of the $b-$tags is missed; 
and the SM single top production. The number of events in the 
signal region 
is expected to be comparable for signal and background~\cite{scott}
for intermediate values of $U_{tc}$. To illustrate this point we plot in 
Fig.~\ref{back} the mass distribution for the top and the charm jet
for $m_{h_t}=200$ and $300$~GeV, $U_{tc}=0.05$, and 
the leading backgrounds~\cite{vecbos} $Wjj$ and $Wb\bar b$, using the  cuts of 
Ref.~\cite{scott}. 
The signal is smeared assuming a resolution given by
$\Delta M_{tj}/M_{tj}=0.80/\sqrt{M_{tj}}$, 
with $M_{tj}$ in GeV.  
For instance, taking a region of $\pm 2\Delta M_{tj}$ around the peaks results in a 
significance $S/\sqrt{B}=10.6$ for $m_{h_t}=200~$GeV, and 
$S/\sqrt{B}=5.8$  for  $m_{h_t}=300~$GeV, for $2fb^{-1}$ of integrated luminosity. 
Although not a substitute for a complete background study, this already
suggests that for rather natural values of $U_{tc}$ the Tevatron in Run~II
will be sensitive to the top-Higgs and top-pions below the $t\bar t$ threshold.

\begin{figure}[h]
\leavevmode
\centering
\epsfig{file=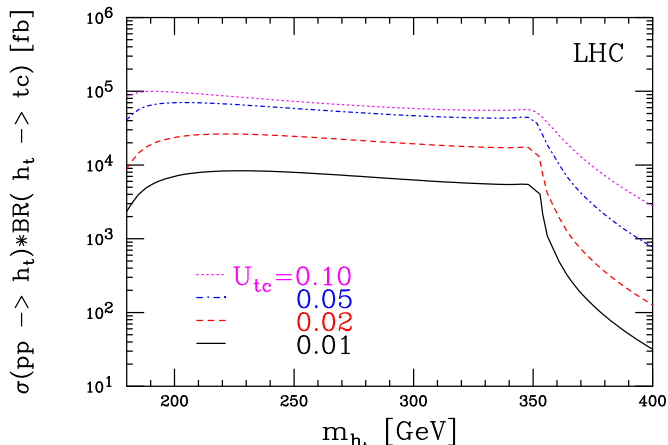,width=6cm,height=8.8cm,angle=90}
\caption{The top-Higgs production cross section via gluon fusion
times its branching ratio to 
$(t\bar c + \bar t c)$ at $\sqrt{s}=14~$TeV, for representative values of the 
parameter $U_{tc}$.
}
\label{lhc}
\end{figure}

At the CERN Large Hadron Collider (LHC) the signal is 
considerably enhanced 
due to the dominance of partonic gluons, and although 
backgrounds can be very large (particularly $t\bar t$ and $Wcj$~\cite{scott})
the signal should be observable in most cases. 
In Fig.~\ref{lhc} we plot the 
top-Higgs production cross section times the branching ratio to top-charm as
a function of $m_{h_t}$ and for various values of the parameter $U_{tc}$. 
For instance, 
for $m_{h_t}=300~$GeV and assuming the same $b$-tagging efficiency as 
for the Tevatron,  as well as the $e,\mu$ decay modes, there will be 
approximately $6000~ {\rm events}/fb^{-1}$.
Thus the discovery of top-Higgses and top-pions at the LHC should be possible 
for a wide range of values of $U_{tc}$.  

There are other models of top-condensation that in principle give rise
to GIM violating interactions and in which the scalar sector 
will also exhibit $tc$ couplings similar to those in Eq.~(\ref{htod}). 
However, they do not always lead to signals as large as the ones 
discussed above. 
An example is the Top see-saw model of Ref.~\cite{dh}. 
Unlike in TATC, here the  Topcolor interactions fully
break the electroweak symmetry implying that $r=1$, and therefore there 
is no 
enhancement in the production via gluon fusion. Moreover, now the 
$WW$ and $ZZ$ dominate the decay of the lightest CP even state, 
analogous to the Higgs, which has a very small branching ratio to 
the flavor changing $tc$ channel.  On the other hand, the low energy
theory of the Top see-saw scenario, 
contains CP-odd composite scalars that can be light enough
to be produced via gluon fusion at the Tevatron and the LHC. 
Although this is not the most general scenario, it is the case for the region 
of parameter space for which the Top see-saw Higgs becomes light~\cite{cdgh}. 
In this case, these pseudo-scalar states could be below the $t\bar t$ threshold
and, since they do not decay into the SM gauge bosons, they may also have a large 
branching fraction to the $tc$ final state.
Due to the suppression at the production end, the discovery of the CP-odd composite 
states of Top see-saw may be very challenging for the Tevatron in Run~II. However, 
it should be a large signal at the LHC.

Finally, we comment on the possibility that the scalar mass, either for 
the top-Higgs or the top-pion, is  above the $t\bar t$ threshold. 
In this case, 
the $t\bar t$ decay mode dominates overwhelmingly, making the flavor changing 
$tc$ mode negligible at the Tevatron. 
The sensitivity of the LHC 
to the $tc$ mode in this case will depend 
critically on the value of $U_{tc}$. 
On the other hand, the effect of the top-condensation scalars on the $t\bar t$ 
cross section
is very small, resulting for instance in a contribution of the order of 
$\delta \sigma(t\bar t)/\sigma(t\bar t)\simeq (1-2)\%$ at  the Tevatron.
 
To summarize, we have seen that the gluon fusion 
production and subsequent flavor changing decay of a top-Higgs or a top-pion 
into the $tc$ final state is likely to be the most important 
signal of top-condensation models at hadron colliders, as long as their masses
are below the $t\bar t$ threshold. This is particularly the case if 
top-gluon masses are larger than $1$~TeV~\cite{cdf}. 
This is a rather model-independent occurrence and it is rooted in the GIM violating
character of the underlying interactions, an essential feature of these models.
The flavor changing decay mode resulting in anomalous single top production has a 
large branching ratio for most of the interesting values of the parameter space.   
Thus the Tevatron with a few $fb^{-1}$ of integrated luminosity
will be sensitive to  Topcolor models, greatly constraining the value of $U_{tc}$. 
Rather small values of $U_{tc}$ can be explored at the LHC, as it can be seen
in Fig.~\ref{lhc}. In both cases a more detailed study of the background
is warranted, in order to establish the sensitivity of each experiment
in the $(U_{tc},m_{h_t})$ parameter space.

The author thanks T. Han, R. Harris, D. Rainwater and especially S. Willenbrock
for useful comments, and the Enrico Fermi Institute
for its  hospitality in the early stages of this work.
This research was supported in part by the U.S.~Department of Energy under  
Grant No.~DE-FG02-95ER40896 and in part by the University of 
Wisconsin Research  
Committee with funds granted by the Wisconsin Alumni Research Foundation.


\end{document}